\documentclass[preprint]{elsarticle}
\usepackage{epsfig}
\usepackage{graphicx}
\usepackage{amssymb}
\newcommand{\AmS}{{\protect\the\textfont2
 A\kern-.1667em\lower.5ex\hbox{M}\kern-.125emS}}
\newcommand{\ba}{\begin{array}}
\newcommand{\ea}{\end{array}}

\def\beq{\begin{equation}}
\def\eeq{\end{equation}}

\def\bea{\begin{eqnarray}}
\def\eea{\end{eqnarray}}
\def\noi{\noindent}
\def\cl{{\rm c}}

\def\beq{\begin{equation}}
\def\eeq{\end{equation}}

\def\bea{\begin{eqnarray}}
\def\eea{\end{eqnarray}}
\def\noi{\noindent}
\def\sl{{\rm s}}
\def\cl{{\rm c}}

\def\ct{\cite}

\def\beq{\begin{equation}}
\def\eeq{\end{equation}}

\def\bea{\begin{eqnarray}}
\def\eea{\end{eqnarray}}

\begin{document}


\title{\Large 
Searching for new physics with three-particle correlations\\
in $pp$ collisions at the LHC
}




\author[cernt,ific]{Miguel-Angel Sanchis-Lozano}
\ead{Miguel.Angel.Sanchis@ific.uv.es}

\author[cerne,uta]{Edward K. Sarkisyan-Grinbaum\corref{cor1}}
\ead{Edward.Sarkisyan-Grinbaum@cern.ch}

\address[cernt]{Theorical Physics Department, CERN, 1211 Geneva 23, 
Switzerland}

\address[ific]{IFIC, Centro Mixto CSIC-Universitat de Val\`encia, Dr. Moliner
50, 46100 Burjassot, Spain}

\address[cerne]{Experimental Physics Department, CERN, 1211 Geneva 23, 
Switzerland}

\address[uta]{Department of Physics, The University of Texas at Arlington,
Arlington, TX
76019, USA}

\cortext[cor1]{Corresponding author}



\begin{abstract}

New phenomena involving pseudorapidity and azimuthal correlations among
final state particles in $pp$ collisions at the LHC can hint at the
existence of hidden sectors beyond the Standard Model. In this paper we
rely on a correlated-cluster picture of multiparticle production, which
was shown to account for the ridge effect, to assess the effect of a
hidden sector on three-particle correlations concluding that there is a
potential signature of new physics that can be directly tested by
experiments using well-known techniques. 

 \end{abstract}

\begin{keyword}
$pp$ interactions at LHC \sep
Models beyond the Standard Model \sep
Multiparticle azimuthal and rapidity correlations \sep
Hidden Valley models \sep
Correlated clusters\\
 \medskip
 {\it Registered preprint number:} arXiv:1802.06703
\end{keyword}





\maketitle




 Multiparticle correlations 
 represent a powerful tool for understanding the underlying 
dynamics of 
particle production mechanisms and to reveal signatures of new 
 and/or unknown phenomena    
 \ct{DeWolf:1995pc,Dremin:2000ep,book,HS:IJMP,CollEff-rev}. 
 Being sensitive to any
 observable 
 deviation from a conventional hadronization process, the 
 correlations are especially suited to search for new physics  
 beyond the Standard Model
 as predicted, e.g., by some Hidden Valley models 
 \cite{Strassler:2006im,Kang:2008ea}.

According to these models, the decay length of hidden particles (e.g. 
hadrons made
of $v$-quarks) can vary wildly, depending on the parameters of the model, leading to completely
distinct phenomenologies. If they are stable, hidden particles will leave
the detector providing a missing energy signature. If, instead, they decay back
into Standard Model particles within the detector, a possible signature 
will
consist of displaced vertices. Finally, 
if hidden particles decay promptly into usual partons, more subtle signatures should
be expected in events generally characterized by 
large multiplicities 
\cite{Strassler:2008fv,Alekhin:2015byh,Knapen:2016hky,Cohen:2017pzm}. 

 In this work, we extend our previous three-particle correlation studies
\ct{Sanchis-Lozano:2017aur} by including a new step in the particle 
production process 
resulting from an additional contribution due to the hypothetical 
formation of an unconventional state of matter on top of the partonic 
cascade 
  as 
  discussed by us earlier \ct{HS:IJMP,Sanchis-Lozano:2015eca}. 
 The study is carried
    out within a model of clusters correlated
    in the collision transverse plane, 
  providing \ct{ridge2part}
  a natural description of the near-side ridge observed in
    two-particle correlations 
for all colliding particles 
and nuclei (for a review, see \ct{CollEff-rev} and similar results
in \ct{Bierlich:2017vhg}). 
    Being generalized to higher-order correlations, the model was found 
\ct{Sanchis-Lozano:2017aur} 
to    show that the ridge-effect should also hold for three-particle
    correlations, in accordance with \ct{Ozonder:2014sra}. 

 The predictions made in this paper can be
 compared with similar
 studies
 at the LHC to search for NP
 expected
 to
modify the parton
shower hadronizing to final-state particles
 \ct{Strassler:2008fv,Nakai:2015swg}.  To this aim,
specific selection cuts should be applied to those
 events
 to be tagged
 as done, e.g.
 in the
 discovery of the nearside ridge in $pp$ interactions.
 In the latter case, the application of
 selection criteria, such as $p_T$ and high multiplicity cuts,
 successfully led to the finding
 of the
 effect.
 Similarly,
 to enhance the NP signal manisfesting through particle correlations,
specific cuts should necessarily be applied to events, such as high-$p_T$
 leptons/photons,
 heavy-flavor-tagging, missing energy, high multiplicity,
 eventually leading to the
 observation of
 structures
 shown in the
 characteristic
 plots
 obtained in
 this paper.

Following the notation of \ct{Sanchis-Lozano:2017aur} though now 
 incorporating a 
hidden sector (HS) contribution, we define one-, two- and three-cluster 
densities, namely,  
 $\rho^{(\cl)}(y_\cl,\phi_\cl; y_{\sl},\phi_\sl)$, 
$\rho_2^{(\cl)}(y_{\cl 1},y_{\cl 2},\phi_{\cl 1},\phi_{\cl 2};y_\sl, \phi_\sl)$ 
and 
$\rho_3^{(\cl)}(y_{\cl 1},y_{\cl 2},y_{\cl 3},\phi_{\cl 1},\phi_{\cl 
2},\phi_{\cl 3};y_\sl,\phi_\sl)$,  
as functions of the cluster 
rapidity $y_\cl$ and azimuthal angle $\phi_\cl$ 
 and  the initial
 hidden particle (source) rapidity $y_\sl$ and azimuthal angle 
$\phi_\sl$.
 The densities 
 satisfy 
 the following conditions:
\[
\int dy_{\cl} d\phi_{\cl}\ \rho^{(\cl)}(y_\cl,\phi_\cl; y_\sl, \phi_\sl)  
=  
\langle 
N_{\cl}^{\sl} 
\rangle \ ,\  
\int  dy_{\cl 1} dy_{\cl 2} d\phi_{\cl 1} d\phi_{\cl 2}\ 
\rho_2^{(\cl)}(y_{\cl 1},y_{\cl 2},\phi_{\cl 1},\phi_{\cl 2}; 
y_\sl, 
\phi_\sl) =
\langle N_{\cl}^{\sl}(N_{\cl}^{\sl}-1) \rangle\ ,
\]
\begin{equation}{\label{eq:rhoc}}
\int  dy_{\cl 1} dy_{\cl 2} dy_{\cl 3}
d\phi_{\cl 1} d\phi_{\cl 2} d\phi_{\cl 3}\ 
\rho_3^{(\cl)}(y_{\cl 1},y_{\cl 2},y_{\cl 3},
\phi_{\cl 1},\phi_{\cl 2},\phi_{\cl 3}; 
 y_\sl, \phi_\sl) 
=
\langle N_{\cl}^{\sl}(N_{\cl}^{\sl}-1)(N_{\cl}^{\sl}-2) \rangle\ ,
\end{equation} 
where $\langle N_{\cl}^{\sl} \rangle$ stands for the average cluster
multiplicity from 
 a HS particle.
 Here and elsewhere in the paper, numerical subscripts for
rapidity and 
azimuthal angle correspond to first, second or third object, 
either particle, cluster or 
 HS particle. On the other hand, $\langle N_{\sl} \rangle$ will denote 
the
average number of HS sources per event, so that the product
$\langle N_{\sl} \rangle \times \langle N_{\cl}^{\sl} \rangle$  gives
the mean number of clusters per collision.

Hereafter, we 
 omit the rapidity variable 
 to focus on the 
 azimuthal dependence.
 To this end, 
 we introduce the production cross section for single 
HS particle production in inelastic hadron collisions as 
 $\rho^{(\sl)}(\phi_\sl) \equiv  (1/\sigma_\sl)\ d\sigma_\sl/d\phi_\sl$, 
 and 
 write for 
 single-particle 
 production:
\begin{equation}{\label{eq:decay1u}}
 \rho(\phi) \equiv \frac{1}{\sigma_{\rm in}}\frac{d\sigma}{d\phi}=
 \int 
  d\phi_{\cl} d\phi_\sl\ 
 \rho^{(\sl)}(\phi_\sl)\ 
 \rho^{(\cl)}(\phi_\cl; \phi_\sl)\ \rho^{(1)}(\phi; \phi_\cl)\ .
\end{equation}

 We introduce the following notation: $\rho^{(\sl)}_2(\phi_{\sl 1}, \phi_{\sl 2}) \equiv (1/\sigma_\sl)\ 
d^2\sigma_\sl/d\phi_{\sl 1} 
d\phi_{\sl 2}$ and 
$\rho^{(\sl)}_3(\phi_{\sl 1}, \phi_{\sl 2}, \phi_{\sl 3}) \equiv 
(1/\sigma_\sl)\ 
d^3\sigma_\sl/d\phi_{\sl 1} d\phi_{\sl 2} d\phi_{\sl 3}$ for 
 double and triple HS production cross sections, respectively; 
$\rho^{(1)}$, $\rho^{(1)}_2$ and $\rho^{(1)}_3$ represent one-, 
two- and three-particle densities from single cluster decay.

Thus we write for the three-particle density
\begin{equation}
 \frac{1}{\sigma_{\rm in}}
 \frac{d^3\sigma} {d\phi_1d\phi_2d\phi_3}\ 
 =\ 
 \int d\phi_{\sl}\  
 \rho^{(\sl)}(\phi_\sl)\
\end{equation}
 \[
\times\
 \left[ 
 \int d\phi_{\cl}\ \rho^{(\cl)}(\phi_{\cl}; \phi_{\sl})\  
\rho^{(1)}_3(\phi_1,\phi_2,\phi_3;\phi_{\cl})+
\int d\phi_{\cl 1} d\phi_{\cl 2}\ \rho^{(\cl)}_2(\phi_{\cl 1},\phi_{\cl 2}; \phi_{\sl})\  
\rho^{(1)}(\phi_1;\phi_{\cl 1})\rho^{(1)}_2(\phi_2,\phi_3;\phi_{\cl 2})\
\right.
\]
\[
\left.
 +\ 
\int d\phi_{\cl 1} d\phi_{\cl 2} d\phi_{\cl 3}\ \rho^{(\cl)}_3(\phi_{\cl 1},\phi_{\cl 2},\phi_{\cl 3}; \phi_{\sl})\  
\rho^{(1)}(\phi_1;\phi_{\cl 1})\rho^{(1)}(\phi_2;\phi_{\cl 2})\rho^{(1)}(\phi_3;\phi_{\cl 3})
 \right]
+\
 \int 
  d\phi_{\sl 1}d\phi_{\sl 2}\ \rho^{(\sl)}_2(\phi_{\sl 1}, \phi_{\sl 2}) 
\
\]
\[
\times\
\left\{
 \left[
\int d\phi_{\cl 1} d\phi_{\cl 2}\ 
\rho^{(\cl)}(\phi_{\cl 1}; \phi_{\sl 1})\ \rho^{(\cl)}(\phi_{\cl 2}; \phi_{\sl 2})\ 
\rho^{(1)}(\phi_1,\phi_{\cl 1})\ \rho^{(1)}_2(\phi_2,\phi_3;\phi_{\cl 2})+\ 
 {\rm combinations}
 \right]
 \right.
\]
\[
\left.
+\int d\phi_{\cl 1} d\phi_{\cl 2} d\phi_{\cl 3}
\left[
\rho^{(\cl)}(\phi_{\cl 1}; \phi_{\sl 1})\ 
\rho^{(\cl)}_2(\phi_{\cl 2},\phi_{\cl 3}; \phi_{\sl 2})\ 
\rho^{(1)}(\phi_1;\phi_{\cl 1})\ \rho^{(1)}(\phi_2;\phi_{\cl 2})\ 
 \rho^{(1)}(\phi_3;\phi_{\cl 3})+\ 
 {\rm combinations}
 \right]
\right\}
\]
\[
 +\
 \int d\phi_{\sl 1} d\phi_{\sl 2} d\phi_{\sl 3}\  
   \rho^{(\sl)}_3(\phi_{\sl 1}, \phi_{\sl 2}, \phi_{\sl 3})\ 
\int d\phi_{\cl 1} d\phi_{\cl 2} d\phi_{\cl 3}
\]
\[
 \times\ 
 \rho^{(\cl)}(\phi_{\cl 1}; \phi_{\sl 1})\ \rho^{(\cl)}(\phi_{\cl 
2}; \phi_{\sl 2})\ 
\rho^{(\cl)}(\phi_{\cl 3}; \phi_{\sl 3})\  
\rho^{(1)}(\phi_1;\phi_{\cl 1})\ \rho^{(1)}(\phi_2;\phi_{\cl 2})\ 
\rho^{(1)}(\phi_3;\phi_{\cl 3})
\ ,
\]
where in the r.h.s., the first line corresponds to the emission 
of secondaries from one and two clusters 
coming from a 
single hidden particle while the second line represents the same but for 
those secondaries from three clusters.
The following two lines correspond
to two and three clusters coming from two
different HS sources. Finally, the 
 last line takes  
into account 
three clusters from three hidden particles.

In order to match our theoretical approach 
to experimental results in terms of 
(pseudo)rapidity and azimuthal differences ($\Delta y_{ij}=y_i-y_j$ and 
$\Delta \phi_{ij}=\phi_i-\phi_j$, $i,j=1,2,3,\ i\neq j$), 
use will be made of integration over Dirac's  $\delta$-functions as 
 in  
 \cite{Sanchis-Lozano:2017aur,ridge2part}. 
Notice that 
only two out of the three rapidity and azimuthal intervals, 
 are independent, chosen here  
  as $\Delta \phi_{12}=\phi_1-\phi_2$
and $\Delta \phi_{13}=\phi_1-\phi_3$. 

Three-particle correlations are thus expressed as a function of the 
rapidity and azimuthal differences
\begin{equation}\label{eq:s3}
s_3(\vec{\Delta y}, \vec{\Delta \phi})\ =\ 
\int d\vec{y}\ d\vec{\phi}\ \vec{\delta}(\Delta y)\ \vec{\delta}(\Delta 
\phi)\ \rho_3(\vec{y},\vec{\phi})\, ,
 \end{equation}
with
\[  
\vec{\Delta y}, \vec{\Delta \phi}\,\;  {\rm  for}\,\; \Delta y_{ij}, 
\Delta 
\phi_{ij}\ ,\,
\vec{y}=(y_1, y_2, y_3)\ ,\ 
\vec{\phi} = (\phi_1, \phi_2, \phi_3)\ ,\ 
d\vec{y}\ d\vec{\phi}\ = 
dy_1dy_2dy_2\ d\phi_1d\phi_2d\phi_3\, ,
 \] 
 \begin{equation}\label{eq:dirac3}
 \vec{\delta}(\Delta y) = \delta(\Delta y_{12}-y_1+y_2)\ 
\delta(\Delta y_{13}-y_1+y_3), \
 \vec{\delta}(\Delta \phi) = \delta(\Delta \phi_{12}-\phi_1+\phi_2)\ 
 \delta(\Delta \phi_{13}-\phi_1+\phi_3)\, . 
 \end{equation}
Here, $\rho_3(\vec{y},\vec{\phi})$ stands for the three-particle case of 
Eq.(\ref{eq:decay1u}), while 
 non-correlated (mixed-event) three-particle distribution
 reads
\begin{equation}\label{eq:b3}
b_3(\vec{\Delta y}, \vec{\Delta \phi})\ =\ 
\int  d\vec{y}\ d\vec{\phi}\ \vec{\delta}(\Delta y)\ \vec{\delta}(\Delta 
\phi)\ \rho(y_1,\phi_1)\ \rho(y_2,\phi_2)\ \rho(y_3,\phi_3)\, , 
\eeq
representing the product of the three single particle distributions.

 In what follows, 
the three-particle 
correlation function, 
\begin{equation}\label{eq:3-cor-short}
c_3(\vec{\Delta y},\vec{\Delta \phi})= \frac{s_3}{b_3}\, ,
 \end{equation}
being  of common use in 
experimental 
 analyses 
 \ct{CollEff-rev}
 is to be compared with our theoretical calculations.

 In the calculations of below, 
 we 
 adopt Gaussian distributions  
in rapidity and azimuthal spaces as usual in cluster models  along with 
the factorization 
hypothesis \cite{Sanchis-Lozano:2017aur,ridge2part}, 
to express production cross sections and 
 cluster densities.
Thus, 
 the single, double and triple HS production cross 
sections 
 read
\[
\frac{1}{\sigma_{\sl}}\frac{d^2\sigma_{\sl}} {dy_\sl d\phi_\sl} \sim  
\exp{\left[-\frac{y_\sl^2}{2\delta_{\sl y}^2}\right]}, \
\frac{1}{\sigma_{\sl}}\frac{d^4\sigma_{\sl}} {dy_{\sl 1} dy_{\sl 2} 
d\phi_{\sl 1} d\phi_{\sl 2}} \sim  
\exp{\left[-\frac{(y_{\sl 1}+y_{\sl 2})^2}{2\delta_{\sl y}^2}\right]}  
\times
\exp{\left[-\frac{(\phi_{\sl 1}-\phi_{\sl 2})^2}{2\delta_{\sl 
\phi}^2}\right]}\, ,
\]
\[
\frac{1}{\sigma_{\sl}}\frac{d^6\sigma_{\sl}} {d\vec{y}_\sl 
d\vec{\phi}_\sl} 
\sim 
\exp{\left[-\frac{(y_{\sl 1}+y_{\sl 2}+y_{\sl 3})^2}{2\delta_{\sl 
y}^2}\right]} 
\times
\left(
\exp{\left[-\frac{(\phi_{\sl 1}-\phi_{\sl 2})^2+(\phi_{\sl 1}-\phi_{\sl 
3})^2+(\phi_{\sl 2}-\phi_{\sl 3})^2}{2\delta_{\sl \phi}^2}\right]}
\right.
\]
 \beq \label{eq:hsgauss}
 \left.
+\
\exp{\left[-\frac{(\phi_{\sl 1}-\phi_{\sl 2})^2}{2\delta_{\sl 
\phi}^2}\right]}
+ \exp{\left[-\frac{(\phi_{\sl 2}-\phi_{\sl 3})^2}{2\delta_{\sl 
\phi}^2}\right]}
+ \exp{\left[-\frac{(\phi_{\sl 1}-\phi_{\sl 3})^2}{2\delta_{\sl 
\phi}^2}\right]}
 \right)\, ,
 \eeq
where $\delta_{\sl y}$ and $\delta_{\sl \phi}$ stand, respectively, for 
the rapidity and 
 azimuthal correlation lengths of the 
 HS particles.

For clusters, one has similarly:
\[
\rho^{(\cl)}(y_\cl,\phi_\cl) \sim  
\exp{\left[-\frac{y_\cl^2}{2\delta_{\cl y}^2}\right]}\, ,\ 
\rho_2^{(\cl)}(y_{\cl 1},\phi_{\cl 1},y_{\cl 2},\phi_{\cl 2}) \sim  
\exp{\left[-\frac{(y_{\cl 1}+y_{\cl 2})^2}{2\delta_{\cl y}^2}\right]}  
\times
\exp{\left[-\frac{(\phi_{\cl 1}-\phi_{\cl 2})^2}{2\delta_{\cl 
\phi}^2}\right]}\, ,
\]
\beq\label{eq:clustercorr}
\rho_3^{(\cl)}(\vec{y_{\cl}},\vec{\phi_{\cl }}) \sim 
\exp{\left[-\frac{(y_{\cl 1}+y_{\cl 2}+y_{\cl 3})^2}{2\delta_{\cl 
y}^2}\right]} 
\times
\left(
\exp{\left[-\frac{(\phi_{\cl 1}-\phi_{\cl 2})^2+(\phi_{\cl 1}-\phi_{\cl 
3})^2+(\phi_{\cl 2}-\phi_{\cl 3})^2
}{2\delta_{\cl \phi}^2}\right]}
\right. 
\eeq
\[
\left.
+\
\exp{\left[-\frac{(\phi_{\cl 1}-\phi_{\cl 2})^2}{2\delta_{\cl 
\phi}^2}\right]}
+ \exp{\left[-\frac{(\phi_{\cl 2}-\phi_{\cl 3})^2}{2\delta_{\cl 
\phi}^2}\right]}
+ \exp{\left[-\frac{(\phi_{\cl 1}-\phi_{\cl 3})^2}{2\delta_{\cl 
\phi}^2}\right]}
\right)\, ,
\]
where $\delta_{\cl y}$ and $\delta_{\cl \phi}$ stand for the rapidity and 
azimuthal cluster correlation lengths, respectively.

It is of paramount importance in our later development to assume that
 $\delta_{\sl \phi} \gg \delta_{\cl \phi}$.
 Indeed, a fast moving cluster should focus particles into a narrow cone
in
the transverse plane (see \cite{Sanchis-Lozano:2017aur}), whereas a
quite massive
hidden source, likely moving at a non-relativistic speed, should spread
out clusters and particles
into a much wider azimuthal angle

 Let us remark that 
 Eqs. (\ref{eq:hsgauss}) and 
(\ref{eq:clustercorr})
can be regarded as parametrizations especially
suitable to 
 model any possible extension of the partonic shower by including 
a new stage on top of it. The 
rapidity Gaussian depending
on the sum of rapidities stems from the
requirement of partial (longitudinal) momentum conservation.
It takes into account different topologies for cluster or hidden particle 
emission once integrated upon their rapidities.
 The azimuthal conditions
are implemented in the Gaussians following
\cite{Sanchis-Lozano:2017aur,ridge2part}, in
order to take into account collinear emission of particles in accordance
with the  ridge effect.

 Since clusters are produced with some non-null (transverse)
momentum, the initial isotropic distribution will be transformed into an
elliptic shape depending on the cluster and emitted particle transverse
velocities. Hence a dependence on the cluster azimuthal angle
$\phi_\cl$ should
remain in the particle densities from single-cluster decay. Then, as
 shown in \ct{ridge2part}, for
these densities,
the rapidity and azimuthal
dependence can be approximately expressed in terms of
Gaussians for highest boosted particles,
i.e.
\[
\rho^{(1)}(y,\phi; y_\cl\phi_\cl) \sim  
\exp{\left[-\frac{(y-y_\cl)^2}{2\delta_{y}^2}\right]}\times 
\exp{\left[-\frac{(\phi-\phi_\cl)^2}{2\delta_{\phi}^2}\right]}\ ,
 \]
\[
\rho^{(1)}_2(y_1,y_2,\phi_1,\phi_2; y_\cl,\phi_\cl) \sim  
\exp{\left[-\frac{(y_1-y_\cl)^2+(y_2-y_\cl)^2}{2\delta_{y}^2}\right]}
\times 
\exp{\left[-\frac{(\phi_2-\phi_\cl)^2+(\phi_2-\phi_\cl)^2}
{2\delta_{\phi}^2}\right]}
 \]
 and similarly for three-particle density.
 The parameter $\delta_y \lesssim 1$ (rapidity units) \ct{Bialas:1973kf}  
is 
usually
 referred to as the cluster decay (pseudo)rapidity ``width'',
while $\delta_{\phi} \simeq 0.14$~radians can be seen 
 the 
cluster decay width 
in the transverse plane \ct{ridge2part}.

Integrating over the phase space of clusters and hidden sources, one 
gets for Eq. (\ref{eq:3-cor-short}),  
 keeping only the
$\Delta \phi_{12}$ and $\Delta \phi_{13}$
 components 
 and  assuming 
 Poisson distribution of clusters and HS particles (see 
\ct{Sanchis-Lozano:2017aur} for 
details of the calculations):

\beq\label{eq:allfunctions}
c_3(\Delta \phi_{12},\Delta \phi_{13})= \frac{1}{\langle N_{\sl}\rangle^2}
h^{(1)}(\Delta 
\phi_{12},\Delta \phi_{13})+
\frac{1}{\langle N_\sl\rangle}h^{(2)}(\Delta \phi_{12},\Delta 
\phi_{13})+h^{(3)}(\Delta 
\phi_{12},\Delta \phi_{13})\ ,
\eeq
where we have fixed the rapidity differences to zero, 
i.e.  $\Delta y_{12}=\Delta y_{13}=0$, thus 
 no 
  explicit 
reference to
rapidity appears 
 in the above expressions. Note that there are the 
weighting factors
as powers of 
$1/{\langle N_{\sl}\rangle}$ have been factorized out 
 as 
 appear in the above expression, while the
$1/{\langle N_{\cl}^{\sl}\rangle}$ factors are included 
in the 
 $h$-functions 
 given below, 
in the limit  $\delta_{\sl \phi}^2 \gg \delta_{\cl\phi}^2\gg 
\delta_{\phi}^2$,
\\

\begin{figure}[t!]
\begin{center}
\includegraphics[scale=0.71]{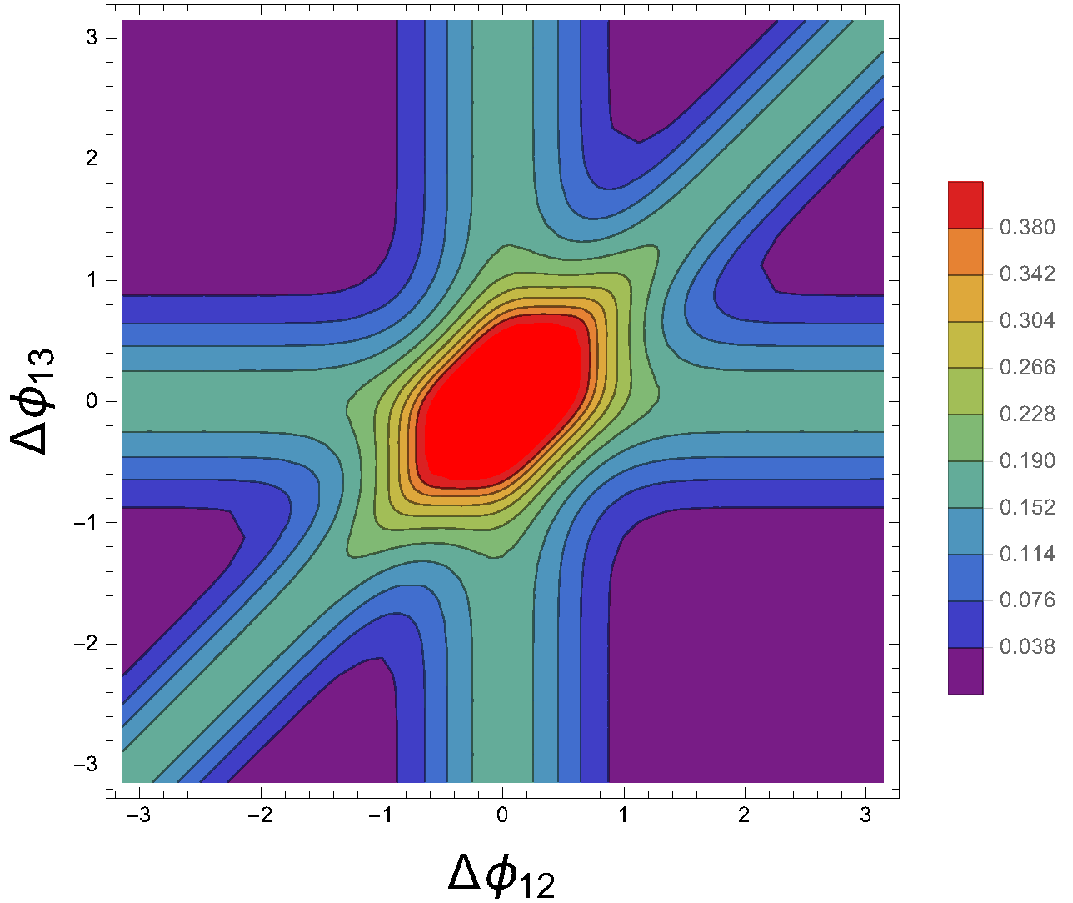}
\hspace{0.5cm}
 \includegraphics[scale=0.71]{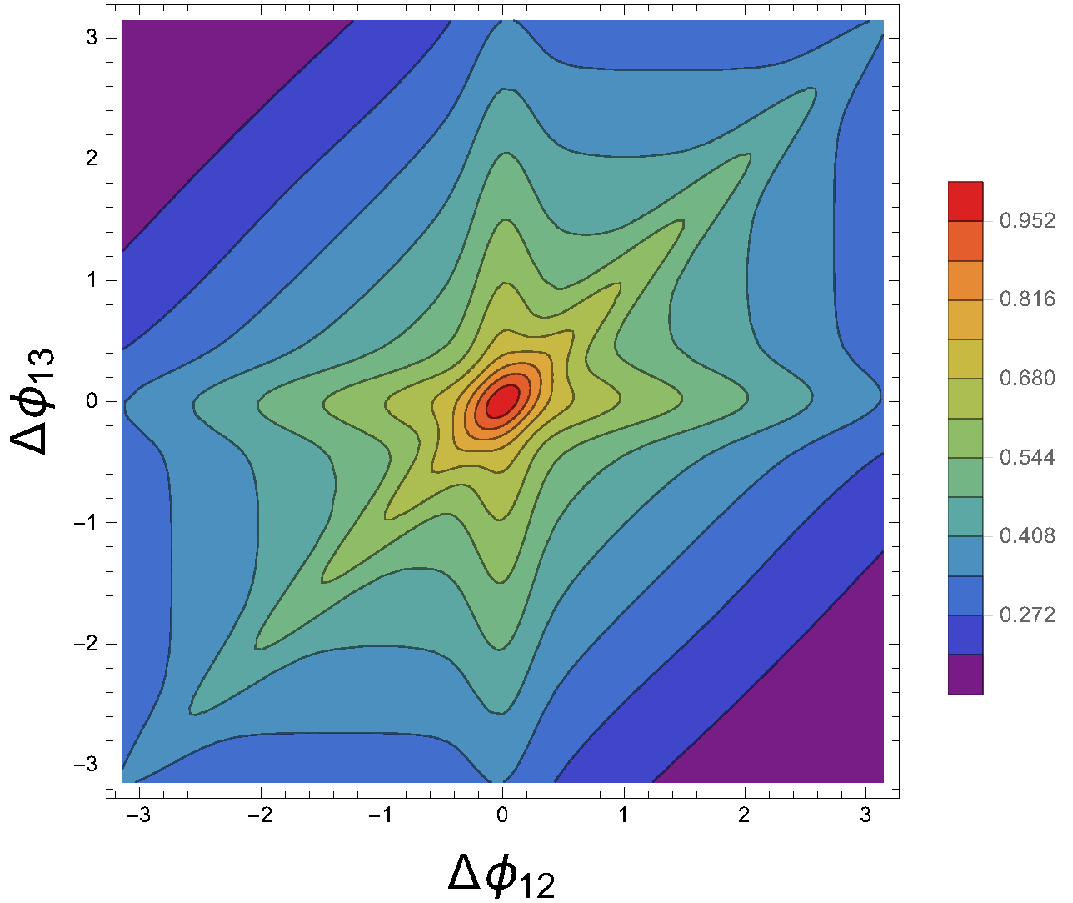}
 \caption{ 
 Contour-plots of $c_3(\Delta \phi_{12},\Delta 
\phi_{13})$ with $\Delta y_{12}=\Delta y_{13}=0$ for
 a two-step cascade (left)
 as found in 
\cite{Sanchis-Lozano:2017aur}, and  
 for
 a three-step
cascade obtained in this work (right). The set of parameters used 
 are given in the text.}
\label{fig:contour}
\end{center}
\end{figure}

\noi
{\it - for single hidden source:}
\beq\label{eq:h1}
h^{\rm (1)}(\Delta \phi_{12},\Delta \phi_{13})\ 
\sim\ 
\frac{1}{\langle N_{\cl}^{\sl}\rangle^ 2}\ \exp{\left[-\frac{(\Delta 
 \phi_{12})^2+(\Delta \phi_{13})^2-\Delta \phi_{12}\Delta 
\phi_{13}}{3\delta_{\phi}^2}\right]}
\eeq
\[
+\
\frac{1}{\langle N_{\cl}^{\sl}\rangle}\ \exp{\left[-\frac{(\Delta 
\phi_{12})^2+(\Delta \phi_{13})^2-\Delta \phi_{12}\Delta 
\phi_{13}}{2\delta_{\cl \phi}^2}\right]}\ 
\]
\[
 \times\ 
 \left(
\exp{\left[-\frac{(\Delta \phi_{12})^2}{4\delta_{\phi}^2}\right]}+
\exp{\left[-\frac{(\Delta \phi_{13})^2}{4\delta_{\phi}^2}\right]}+  
\exp{\left[-\frac{(\Delta \phi_{12})^2+(\Delta \phi_{13})^2-2\Delta 
\phi_{12}\Delta \phi_{13}}{4\delta_{\phi}^2}\right]}
 \right)
\]
\[
+\
\exp{\left[-\frac{(\Delta \phi_{12})^2+(\Delta \phi_{13})^2-\Delta 
\phi_{12})\Delta \phi_{13})}{3\delta_{\cl \phi}^2}\right]}
\]
\[
+\
\exp{\left[-\frac{(\Delta 
\phi_{12})^2}{2\delta_{\cl \phi}^2}\right]}+\exp{\left[-\frac{(\Delta 
\phi_{12})^2}
{2\delta_{\cl \phi}^2}\right]}+ 
\exp{\left[-\frac{(\Delta \phi_{12})^2+(\Delta \phi_{13})^2-2\Delta 
 \phi_{12}\Delta \phi_{13}}{2\delta_{\cl \phi}^2}\right]}
\]

\noi
{\it - for two hidden sources:}
\[
h^{\rm (2)}(\Delta \phi_{12},\Delta \phi_{13})\ \sim\   
 \left(
\frac{1}{\langle N_{\cl}^{\sl}\rangle} 
\exp{\left[-\frac{(\Delta \phi_{12})^2+(\Delta \phi_{13})^2-\Delta 
\phi_{12}\Delta \phi_{13}}
 {2\delta_{\cl \phi}^2+\delta_{\sl \phi}^2}\right]}\right.
\]
\beq\label{eq:h2} 
\left.
+\ \exp{\left[-\frac{(\Delta \phi_{12})^2+(\Delta \phi_{13})^2-\Delta 
\phi_{12}\Delta \phi_{13}} {3\delta_{\cl \phi}^2+2\delta_{\sl \phi}^2}\right]}
\right)
\eeq
\[
\times\
\left(
\exp{\left[-\frac{(\Delta \phi_{12})^2}{4\delta_{\phi}^2}\right]}+
\exp{\left[-\frac{(\Delta \phi_{13})^2}{4\delta_{\phi}^2}\right]}+ 
\exp{\left[-\frac{(\Delta \phi_{12})^2+(\Delta \phi_{13})^2-2\Delta 
\phi_{12}\Delta \phi_{13}}{4\delta_{\phi}^2}\right]}
\right)
\]
 
\noi
{\it - for three hidden sources:}
\beq\label{eq:h3}
h^{\rm (3)}(\Delta \phi_{12},\Delta \phi_{13})\ \sim\  
\exp{\left[-\frac{(\Delta \phi_{12})^2+(\Delta \phi_{13})^2-\Delta 
\phi_{12}\Delta \phi_{13}}
 {3\delta_{\cl \phi}^2+\delta_{\sl \phi}^2}\right]}
\eeq
\[
+\
\exp{\left[-\frac{(\Delta \phi_{12})^2}{2(2\delta_{\cl \phi}^2+\delta_{\sl \phi}^2)}\right]}+
\exp{\left[-\frac{(\Delta \phi_{13})^2}{2(2\delta_{\cl \phi}^2+\delta_{\sl \phi}^2)}\right]}+ 
\exp{\left[-\frac{(\Delta \phi_{12})^2+(\Delta \phi_{13})^2-2\Delta 
\phi_{12}\Delta \phi_{13}}{2(2\delta_{\cl 
\phi}^2+\delta_{\sl \phi}^2)}\right]}\, .
\]
\\
Each term in the above expressions can be put in correspondence with another one
from the set of Eqs.(3). In fact, Eqs.(\ref{eq:h1})-(\ref{eq:h3}) 
represent
a generalization of the equivalent expressions in our previous work
 \cite{Sanchis-Lozano:2017aur} once a hidden sector is included. 
 Notice also that
 the key point for the physical consequences to be explored below
is {\em not} having three (or more) steps of clustering, instead of two,
but
the fact that the first cluster provides a long-range correlation length
throughout
the whole chain of subsequent clusters and final particles.

 The behaviour of the three-particle 
correlation function
$c_3(\Delta \phi_{12},\Delta \phi_{13})$
as a function of the azimuthal differences $\Delta \phi_{12}$ and 
 $\Delta \phi_{13}$ (for $\Delta y_{12}=\Delta y_{13}=0$), 
as it could be 
 measured experimentally, is shown in Fig. \ref{fig:contour}. 
 The left panel  shows 
 the contour-plot of
 the $c_3$-function 
 corresponding to a (two-step) standard cascade as obtained earlier in  
 \cite{Sanchis-Lozano:2017aur}. 
 The right panel 
 shows 
 the new result corresponding to a three-step cascade that we identify
with the possible existence of a new stage of matter on top of the conventional
parton shower yielding final state particles. We tentatively 
set $\langle N_{\sl}\rangle=2$, $\langle N_{\cl}^{\sl}\rangle=3$ for
the average multiplicities. 
On the other hand, the pressumably large mass of HS particles
implies that their velocities should be considerably
smaller than those of clusters and final final state particles. 
Thereby we choose
$\delta_{\sl \phi} \simeq \pi$ as a reference value for the correlation length 
in the transverse plane between 
HS particles. Besides, there is an overall normalization 
of the $c_3(\Delta \phi_{12},\Delta \phi_{13})$ function to unity 
at $\Delta \phi_{12}=\Delta \phi_{13}=0$. 
It is worthwhile remarking too that
the main features of the right-hand plot in Fig.~\ref{fig:contour} remain 
 almost unchanged
 under
reasonable variations of the above-mentioned parameters.

It is not difficult to understand the underlying reason for such 
different behaviours. 
Long-range correlations of final-state particles 
are inherited from a hidden source convoluting with the shorter correlations from
clusters, thereby stretching the ``radii'' of the ``spiderweb-type'' 
structure.

On the other hand, the two-dimensional plot corresponding to (pseudo)rapidity 
 intervals 
remains 
practically 
the same (thus not shown in this paper, see \cite{Sanchis-Lozano:2017aur}).
This can be attributed to the fact that
 long rapidity correlations are already present in the conventional
cascade so that an additional HS source in the partonic shower
with $\delta_{\sl y} \simeq \delta_{\cl y}$ does not significantly alter 
the plot.

\begin{figure}[t!]
\begin{center}
\includegraphics[scale=0.6]{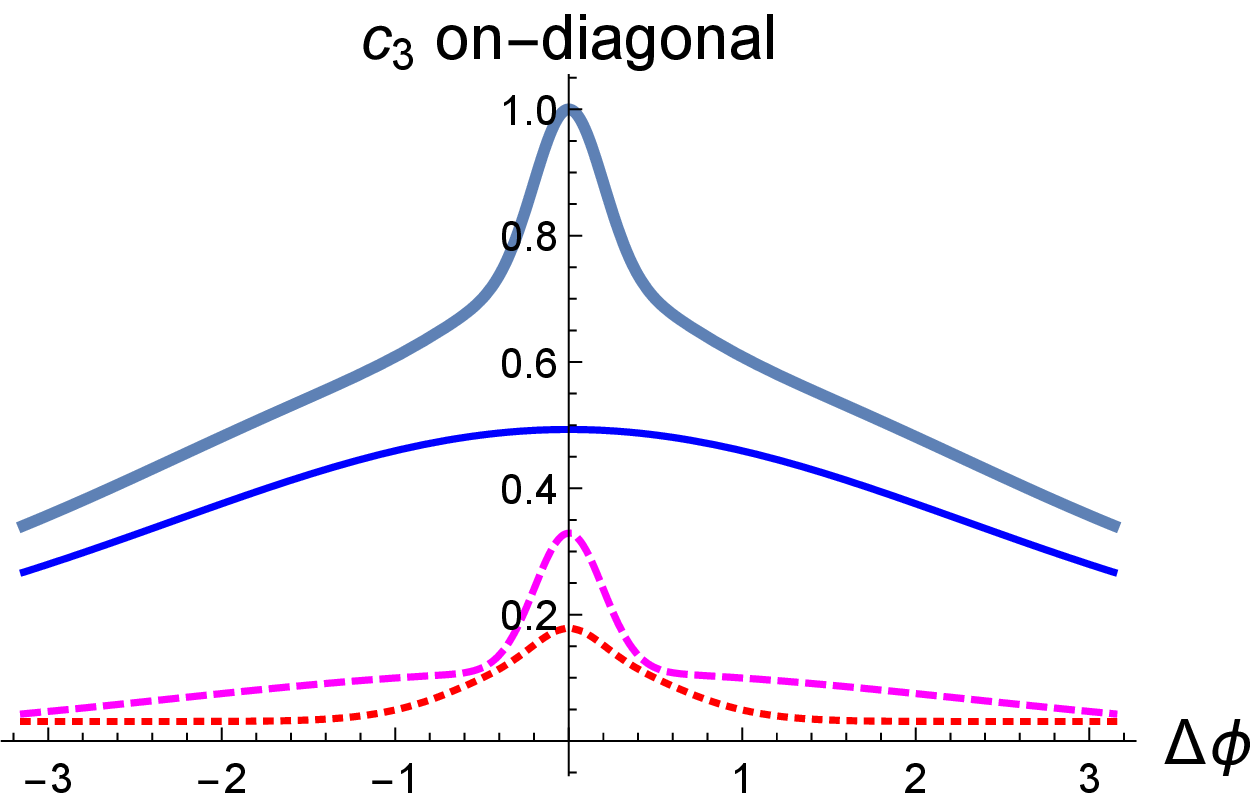}
\hspace{0.5cm}
\includegraphics[scale=0.6]{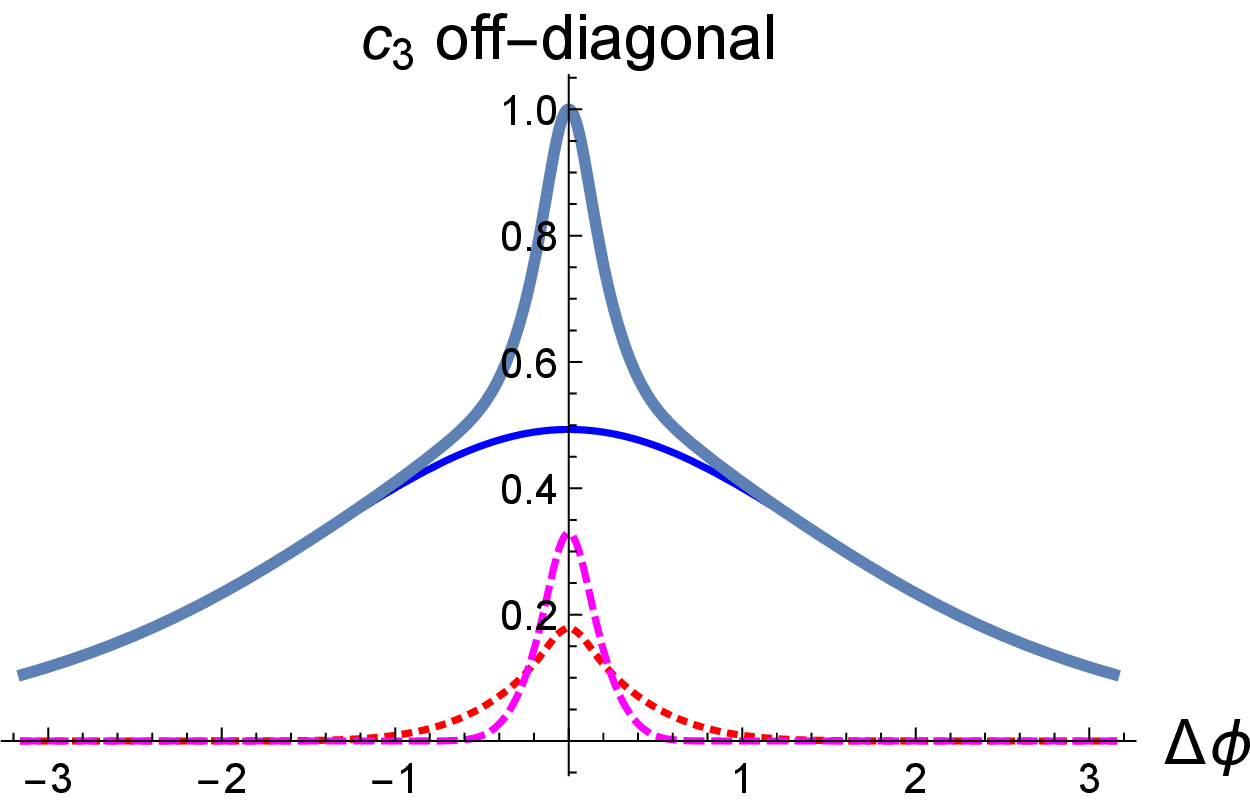}
\caption{Diagonal (left) and off-digonal (right) 
projections of the azimuthal contour plot of $c_3(\Delta \phi_{12},\Delta 
\phi_{13})$ 
 from Fig. \ref{fig:contour}
 for
a three-step cascade
  as obtained in this work. 
The dotted (red), dashed (magenta) and thin solid (blue) curves show 
the weighted 
contributions from one, two and three 
hidden particles, respectively, and 
the
 thick (turquoise) curve
shows the sum of these contributions.}
\label{fig:projphi}
\end{center}
\end{figure}

 Figure \ref{fig:projphi} shows the projection plots of the 
 $c_3$ function  
 along the diagonal 
($\Delta \phi_{12}=\Delta \phi_{13}$, left panel) and 
 off the diagonal ($\Delta \phi_{12}=-\Delta \phi_{13}$, right panel)
 under the $\Delta y_{12}=\Delta y_{13}=0$ condition. 
 A different behaviour can again be remarked in both plots, as the 
on-diagonal correlation length
is appreciably longer than the off-diagonal correlation length. 
The contributions from the different
pieces $h^{\rm (1)}$, $h^{\rm (2)}$ and $h^{\rm (3)}$ are 
also separately shown.  Let us observe that the contribution from the $h^{(3)}$ piece is 
mainly responsible of the  ``web'' structure in the plot.

Summarizing, in this work 
a potential signature of new physics
 is shown to be observed  
 in three-particle azimuthal correlations
 which can be 
directly 
tested in experiments at the LHC. Our results can be extended
to other than $pp$ collisions. 
According to our study, the effect of a new stage of matter,
as considered here, would manifest as
 a ``web'' structure in the 
 three-particle two-dimensional correlation plot in azimuthal space.
 Such a signature should be considered as complementary to
other possible signatures, helpful to discriminate among 
 distinct phenomenologies 
 from Hidden Valley
 models.

\subsubsection*{Acknowledgements}

This work has been partially supported by the Spanish MINECO under grants 
FPA2014-54459-P and FPA2017-84543-P, by the Severo Ochoa Excellence 
Program under grant SEV-2014-0398 and by the Generalitat Valenciana under 
grant GVPROMETEOII 2014-049. M.A.S.L. thanks the CERN Theoretical Physics 
Department, where this work has been done, for its warm hospitality.
\\

\end{document}